\documentclass[prb,twocolumn,preprintnumbers,amsmath,amssymb,showpacs,nofootinbib,floatfix]{revtex4}

\usepackage{graphicx,bm}

\newcommand{\nuxy}{{\nu_{\textsc{xy}}}}
\newcommand{\oxy}{{\omega_{\textsc{xy}}}}

\newcommand{\exy}{{\eta_{\textsc{xy}}}}
\newcommand{\ou}{{\omega_{u}}}

\makeatletter
\def\graphicscale{\twocolumn@sw{0.3}{0.4}}
\def\graphicthreescale{\twocolumn@sw{0.3}{0.4}}

\begin{document}

\title{Bose-Einstein condensation and critical behavior of
  two-component bosonic gases}

\author{Giacomo Ceccarelli,$^1$ Jacopo Nespolo,$^1$ 
Andrea Pelissetto,$^2$ and Ettore Vicari$^1$} 

\address{$^1$ Dipartimento di Fisica dell'Universit\`a di Pisa and
  INFN, Largo Pontecorvo 3, I-56127 Pisa, Italy} \address{$^2$
  Dipartimento di Fisica dell'Universit\`a di Roma ``La Sapienza''
  and INFN, Sezione di Roma I, I-00185 Roma, Italy}

\date{\today}

\begin{abstract}

We study Bose-Einstein condensation (BEC) in three-dimensional
two-component bosonic gases, characterizing the universal behaviors of
the critical modes arising at the BEC transitions. For this purpose,
we use field-theoretical (FT) renormalization-group (RG) methods and
perform mean-field and numerical calculations.  The FT RG analysis is
based on the Landau-Ginzburg-Wilson $\Phi^4$ theory with two complex
scalar fields which has the same symmetry as the bosonic system. In
particular, for identical bosons with exchange $\mathbb{Z}_{2}$
symmetry, coupled by effective density-density interactions, the
global symmetry is ${\mathbb Z}_{2,e} \otimes {\rm U}(1) \otimes {\rm
  U}(1)$.  At the BEC transition it may break into ${\mathbb Z}_{2,e}
\otimes {\mathbb Z}_2 \otimes {\mathbb Z}_2$ when both components
condense simultaneously, or to ${\rm U}(1) \otimes {\mathbb Z}_2$ when
only one component condenses. This implies different universality
classes for the corresponding critical behaviors.  Numerical
simulations of the two-component Bose-Hubbard model in the hard-core
limit support the RG prediction: when both components condense
simultaneously, the critical behavior is controlled by a decoupled XY
fixed point, with unusual slowly-decaying scaling corrections arising
from the on-site inter-species interaction.

\end{abstract}


\maketitle



\section{Introduction}
\label{intro}

Experiments with cold atoms\cite{CW-02,Ketterle-02,BDZ-08} have
provided the opportunity to investigate Bose-Einstein condensation
(BEC) in dilute interacting atomic gases.  In the BEC a macroscopic
number of bosonic atoms, the so-called condensate, occupy the
lowest-energy quantum state at a finite temperature.  The phase of the
condensate wave function provides the order parameter at the
transition.  BEC transitions are generically expected to belong to the
three-dimensional (3D) XY universality class, which is characterized
by the spontaneous breaking of an Abelian U(1) symmetry.  The same
universal critical behavior is observed in the superfluid transition
in $^4$He,~\cite{Lipa-etal-96,CHPV-06} in transitions characterized by
density or spin waves (as it occurs in some liquid crystals), in
magnetic systems with easy-plane anisotropy, etc.~\cite{PV-02} The XY
behavior at the 3D BEC transition has been supported by experimental
measurements of the diverging correlation length in a cold-atom
bosonic gas.~\cite{DRBOKS-07} Cold-atom experiments have been extended
to mixtures of homonuclear and heteronuclear bosonic
gases,\cite{PhysRevLett.78.586,
  PhysRevLett.81.1539,PhysRevLett.82.2228,PhysRevA.63.051602,
  PhysRevLett.99.190402,
  PhysRevLett.103.245301,PhysRevLett.105.045303,
  PhysRevA.80.023603,PhysRevA.82.033609,Nature.396.345,
  PhysRevLett.85.2413,PhysRevLett.89.053202,
  PhysRevLett.89.190404,PhysRevLett.99.010403,PhysRevA.77.011603,
  PhysRevLett.100.210402,PhysRevLett.101.040402,
  PhysRevA.79.021601,PhysRevA.84.011610} which also show BEC
phenomena.  Several theoretical studies have discussed various aspects
of the behavior of mixtures of bosonics gases, see, e.g.,
Refs.~\onlinecite{HS-96,Boninsegni-01,AHDL-03,DDL-03,KS-03,PC-03,
  KG-04,KPS-04,ICSG-05,PSP-08,SCPS-09,HSH-09,CSPS-10,
  FHRSB-11,Pollet-12,ACV-14,LCD-14,GBS-15}, such as low-dimensional
behaviors, magnetic-like behaviors at the $n=1$ Mott phases, etc.
However, some issues call for further investigations, such
as the critical behaviors at the finite-temperature
normal-to-superfluid transitions which arise from different BEC
patterns.

In this paper we study BEC in mixture of 3D bosonic gases, focussing
on the critical behaviors at the finite-temperature transitions
arising from BEC.  In particular, we consider a system of two
identical boson gases with density-density interactions.
Equivalently, we may interpret this system as made up by a single
two-component boson gas.  An example is provided by the lattice
two-component Bose-Hubbard (2BH) model
\begin{eqnarray}
H &=& - t \sum_{s} \sum_{\langle {\bm x \bm y} \rangle} 
(b_{s\bm{x}}^\dagger b_{s\bm{y}}^{} + {\rm h.c}) -  
\mu \sum_{s\bm{x}} n_{s\bm{x}} 
\label{sBH}\\
&+& \frac{1}{2} V
 \sum_{s\bm{x}} n_{s\bm{x}} (n_{s\bm{x}}-1)  + 
U \sum_{\bm{x}} n_{1\bm{x}} n_{2\bm{x}},
\nonumber
\end{eqnarray}
where $\langle {\bm x \bm y} \rangle$ indicates the nearest-neighbor
sites of a cubic lattice, the subscript $s$ labels the two species,
and $n_{s\bm{x}}\equiv b_{s\bm{x}}^\dagger b_{s\bm{x}}^{}$ is the
density operator. The Hamiltonian is symmetric under U(1)
transformations acting independently on the two species and under the
${\mathbb Z}_{2}$ transformation exchanging the two bosons.  The
two-component boson gas shows a quite complex phase diagram in the
space of the model parameters, i.e., the temperature $T$, the chemical
potential $\mu$, and the on-site couplings $U$ and $V$.  In the
following we set $t=1$ for the hopping parameter without loss of
generality.

We investigate the critical behavior of systems like the 2BH model by
field-theoretical (FT) renormalization-group (RG) methods, mean-field
and numerical approaches.  We show that transitions in these
two-component systems may be associated with different spontaneous
breakings of the global symmetry
\begin{equation}
  {\mathbb Z}_{2,e} \otimes {\rm U}(1) \otimes {\rm U}(1).
\label{symm}
\end{equation}
This symmetry may break to ${\mathbb Z}_{2,e} \otimes {\mathbb Z}_2
\otimes {\mathbb Z}_2$ when both components condense simultaneously,
or to ${\rm U}(1) \otimes {\mathbb Z}_2$ when only one component
condenses, with two different universality classes for the
corresponding critical behaviors.

When both components condense simultaneously, the RG analysis shows
that the critical behavior is controlled by a decoupled 3D XY fixed
point (FP).  Thus, the transition belongs to the 3D XY universality
class associated with the symmetry breaking U(1)$\to$${\mathbb
  Z}_2$. However, the irrelevant density-density interaction between
the two components gives rise to scaling corrections that decay very
slowly, as $\xi^{\,-0.022}$, where $\xi$ is the diverging length scale
at the transition. Such scaling corrections are not present in
standard transitions belonging to the XY universality class, such as
at the BEC transition of a single bosonic
species.\cite{CN-14,CTV-13,CR-12,CPS-07} In that case, scaling
corrections decrease significantly faster, as $\xi^{\,-0.78}$. If,
instead, only one component condenses, the RG analysis predicts a
different critical behavior, which belongs to the same universality
class as that of the continuous transitions in chiral models with
O$(2)\otimes$O$(2)$ symmetry.\cite{Kawamura-88} We also present a RG
analysis of the behavior of mixtures of nonidentical bosons with
density-density interactions. In this case the phase diagram is
characterized by the presence of bicritical or tetracritical points
where various transition lines meet.

The paper is organized as follows.  In Sec.~\ref{ftrgan} we present
the RG analysis of the Landau-Ginzburg-Wilson (LGW) theory, which is
expected to describe the critical transitions in two-component bosonic
systems with density-density interactions.  In Sec.~\ref{mfan} we
discuss the phase diagram of the 2BH model (\ref{sBH}) in the
mean-field approximation.  Sec.~\ref{qmc} is devoted to a numerical
study of the 2BH model in the hard-core, $V\to\infty$, limit. At the
transition both components condense. We show that the results can be
explained by a 3D XY critical behavior with slowly-decaying scaling
corrections, as predicted by the RG analysis.  Finally, in
Sec.~\ref{conclu} we draw our conclusions.

\section{Field-theoretical renormalization-group analysis}
\label{ftrgan}

We wish now to classify the finite-temperature transitions in the
phase diagram of systems consisting of two identical boson species
with density-density interactions. For this purpose, we study the RG
flow of the effective LGW $\Phi^4$ theory associated with the critical
modes.\cite{Aharony-76,ZJ-book,PV-02,v-07} Within the FT RG approach,
one first identifies the order parameter. Then, one considers the
$\Phi^4$ Hamiltonian with the most general fourth-order potential in
the order parameter that has the same symmetry properties as the
original system.  The possible critical behaviors are determined by
the stable FPs of the RG flow.  Each of them corresponds to a
different universality class, associated with the symmetry breaking
that occurs in the parameter region in which the FP is located. The
stable FPs determine the universal scaling properties, such as the
critical exponents, the scaling functions, etc. Note that only systems
which are in the attraction domain of the stable FPs undergo
continuous transitions.  Systems corresponding to LGW theories with
parameters that are outside the FP attraction domains, or that belong
to the instability region, are predicted to undergo first-order
transitions.

At a BEC transition, the condensate behaves like the magnetization in
magnetic systems, i.e., $\langle b_{s{\bm x}} \rangle \sim
(T_c-T)^\beta$ as $T_c$ is approached from below. Critical modes
develop a diverging length scale $\xi \sim |T-T_c|^{-\nu}$, while the
two-point function at the critical point decays algebraically
as\cite{ZJ-book,PV-02} $G({\bm x})\sim |{\bm x}|^{-1-\eta}$.  The
exponents $\beta$, $\nu$, and $\eta$ are universal---they only depend
on the universality class---and are related by the scaling relation
$\beta=\nu (1+\eta)/2$.

\subsection{LGW theory for two-component boson gases}
\label{ftrgantwof}

In the case of a mixture of bosonic gases, we associate a complex
field $\varphi_{s}({\bm x})$, $s=1,2$, with each bosonic species.
Since we consider finite-temperature transitions of 3D quantum
systems, we must consider a three-dimensional LGW model. As mentioned
in the introduction, the relevant symmetry of the systems we consider,
such as the 2BH model (\ref{sBH}), is ${\mathbb Z}_{2,e} \otimes {\rm
  U}(1) \otimes {\rm U}(1)$.  The Hamiltonian is therefore
\begin{eqnarray}
{\cal H}_{\rm LGW} &=& \int d^3x\,\Bigl[
\sum_{s,\mu} |\partial_\mu \varphi_s({\bm x})|^2  +  
r \sum_s |\varphi_s({\bm x})|^2 \label{phi4s}\\
&+& g \sum_s |\varphi_s({\bm x})|^4 
+ 2 u\, |\varphi_1({\bm x})|^2 |\varphi_2({\bm x})|^2\Bigr],
\nonumber
\end{eqnarray}
where the potential is the most general one under symmetry
(\ref{symm}).  The Hamiltonian is bounded from below for $g>0$ and
$g+u>0$.  The quartic couplings $g$ and $u$ are related to the
intra-species $V$ and inter-species $U$ on-site couplings of the 2BH
model (\ref{sBH}). In particular, $u$ must vanish when $U$ vanishes,
leaving two decoupled LGW theories, one for each bosonic gas.

General information on the phase diagram of model (\ref{phi4s}) can be
inferred by a straightforward mean-field analysis, e.g., by
determining the minima of the potential
\begin{eqnarray}
V(\varphi) &=& r \sum_s |\varphi_s|^2  +
g \sum_s |\varphi_s|^4  
+ 2 u |\varphi_1|^2 |\varphi_2|^2.  \label{potential}
\end{eqnarray}
For $r > 0$ the potential is minimized by $\varphi_s = 0$, while for
$r<0$ the minimum depends on the sign of $w\equiv g-u$.  If $w>0$, the
minimum occurs when both field components condense, i.e., for $\langle
\varphi_1 \rangle = \langle \varphi_2 \rangle\neq 0$.  This implies
the symmetry-breaking pattern
\begin{equation}
  {\mathbb Z}_{2,e} \otimes {\rm U}(1) \otimes {\rm U}(1)
  \to {\mathbb Z}_{2,e}
  \otimes {\mathbb Z}_2 \otimes {\mathbb Z}_2,
\label{symmp1}
\end{equation}
i.e., each U(1) group breaks into ${\mathbb Z}_2$.  Instead for $w<0$,
we have $\langle \varphi_1 \rangle \neq 0$ and $\langle \varphi_2
\rangle=0$, or viceversa.  Thus, the exchange symmetry and only one of
the two U(1) groups are broken, so that
\begin{equation}
  {\mathbb Z}_{2,e} \otimes {\rm U}(1) \otimes 
{\rm U}(1)
  \to {\rm U}(1) \otimes {\mathbb Z}_2. 
\label{symmp2}
\end{equation}
On the boundary line $w=0$, the LGW theory (\ref{phi4s}) is equivalent
to the O(4) vector model.

\subsection{RG flow and critical behaviors}
\label{critbeh}

\begin{figure}
\vskip5mm
\includegraphics[width=4.2cm]{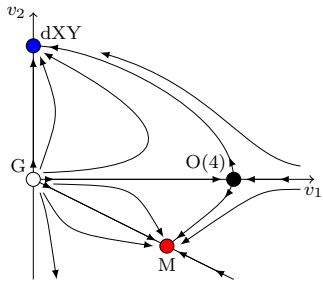}
\caption{Sketch of the RG flow of the $MN$ model (\ref{HMN}) for
  $M=N=2$.  The relevant stable FPs are the decoupled XY FP (dXY),
  corresponding to FP (iii) in the list reported in the text, and FP
  $M$, corresponding to FP (iv).  }
\label{flux-MN}
\end{figure}

The LGW theory (\ref{phi4s}) is a particular case of the so-called
$MN$ model~\cite{Aharony-76,PV-02}
\begin{eqnarray}
\int d^3 x\Bigl[  \sum_{ai} (\partial_\mu \phi_{ai})^2 + r \phi_{ai}^2   
+ \sum_{ijab} (v_1 + v_2 \delta_{ij}) \phi_{ai}^2 \phi^2_{bj}\Bigr],
\label{HMN}
\end{eqnarray}
where $\phi_{ai}$ is an $M\times N$ matrix, i.e., $a=1,\ldots,N$ and
$i=1,\ldots,M$.  Indeed, Hamiltonian (\ref{phi4s}) reduces to
(\ref{HMN}) for $M=N=2$, if we set $\phi_{1i}={\rm Re}\,\varphi_i$,
$\phi_{2i} = {\rm Im} \,\varphi_i$, and
\begin{equation}
g = v_{1} + v_{2},\qquad u = v_{1}, \qquad w \equiv g-u= v_2.
\label{uvcorr}
\end{equation}
The RG flow of the $MN$ models has been studied by various FT
methods.\cite{Aharony-76,PV-02,PV-05} A sketch of the RG flow in the
case $M=N=2$ is shown Fig.~\ref{flux-MN}.  There are several FPs in
the plane of the renormalized quartic couplings $v_{1}$ and
$v_{2}$:~\cite{PV-05} (i) the trivial Gaussian FP for $v_{1}=v_{2}=0$
which is unstable against both quartic perturbations present in
Hamiltonian (\ref{HMN}); (ii) the O(4)-symmetric FP for $v_{2}=0$ and
$v_{1}>0$ which is unstable with respect to the quartic term
proportional to $v_{2}$ in Eq.~(\ref{HMN}); (iii) a stable decoupled
XY FP with $v_{1} = 0$ and $v_{2}>0$, with attraction domain in the
region $v_{2}>0$; (iv) a stable FP for $v_{2}<0$ with attraction
domain in the region $v_{2}<0$.

\begin{figure}
\vskip5mm
\includegraphics[width=4.2cm]{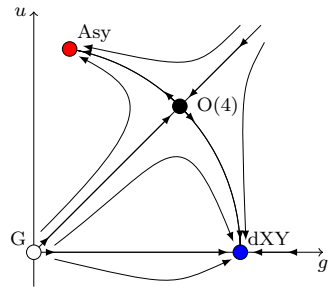}
\caption{Sketch of the  RG flow of the LGW theory for a mixture of
  two identical bosonic gases. The relevant stable FPs are the
  decoupled XY FP (dXY), which controls the RG flux for $g-u > 0$ (in
  this case the exchange symmetry is conserved), and a second FP
  (Asy), relevant for $g - u < 0$ (the exchange symmetry is broken).
}
\label{flux-LGW}
\end{figure}

It is also possible to map the LGW theory (\ref{phi4s}) onto the 
chiral LGW theory with O($N)\otimes$O($M$) symmetry defined by
\cite{Kawamura-88,CPPV-04,PRV-01}
\begin{eqnarray}
\int d^3 x &&
\Bigl\{  \sum_{ai} [(\partial_\mu \phi_{ai})^2 + r \phi_{ai}^2]
+ {(u_0-v_0)} \sum_{aij} \phi^2_{ai} \phi^2_{aj} + \nonumber  \\
&&
  {v_0} \sum_{abij} \phi_{ai} \phi_{bi}\phi_{aj} \phi_{bj}
\Bigr\},
\end{eqnarray}
where $\phi_{ai}$ is an $M\times N$ matrix. The two models are equivalent 
for $M = N=2$, if we identify fields and couplings as follows:
\begin{eqnarray}
\phi_{11} &=& {1\over \sqrt{2}} ({\rm Re}\, \varphi_1 - {\rm Im}\, \varphi_2),
\nonumber \\
\phi_{12} &=& {1\over \sqrt{2}} ({\rm Im}\, \varphi_1 - {\rm Re}\, \varphi_2),
\nonumber \\
\phi_{21} &=& {1\over \sqrt{2}} ({\rm Im}\, \varphi_1 + {\rm Re}\, \varphi_2),
\nonumber \\
\phi_{22} &=& {1\over \sqrt{2}} ({\rm Re}\, \varphi_1 + {\rm Im}\, \varphi_2),
\end{eqnarray}
and $g = u_0 - v_0/2$, $u = u_0 + v_0/2$. Note that the couplings of
the chiral and of the $MN$ model are related by $v_0 = - v_2$ and $u_0
= v_1 + v_2/2$, so that the FP (iv) mentioned above corresponds to the
chiral FP that has been extensively discussed in
Refs.~\onlinecite{Kawamura-88,PRV-01,CPPV-04,foot-controversy}.

According to the correspondence (\ref{uvcorr}), the equivalent model
(\ref{phi4s}) presents two stable FPs with attraction domains
separated by the line $w=g-u=0$, along which the unstable
O(4)-symmetric FP is located.
The corresponding
RG flows are  sketched in  Fig.~\ref{flux-LGW}.

If $w>0$ the finite-temperature transition is characterized by the
simultaneous condensation of both species. The stable FP which
determines the critical behavior is located along the line $u=0$, thus
representing two decoupled U(1)-symmetric models.  This implies that
it belongs to the 3D XY universality class, whose critical exponents
are~\cite{CHPV-06} $\nuxy=0.6717(1)$ and $\exy=0.0381(2)$.  However,
scaling corrections decay much slower than in the U(1)-symmetric
theory with a single complex field, where the leading irrelevant
perturbation has RG dimension $y_g \equiv -\oxy =
-0.785(20)$.\cite{CHPV-06,PV-02} This is related to the RG dimension
at the decoupled XY FP of the interaction operator between the two
complex order parameters.  Since the RG dimension of the energy
operator $\int d^3x\, \phi^2$ is $1/\nuxy$, the RG dimension of the
interspecies interaction operator $\int d^3x\, |\varphi_1|^2
|\varphi_2|^2$ is
\begin{equation}
y_u = 2/\nuxy - 3 = - 0.0225(4).
\label{ypder}
\end{equation}
Since $y_u<0$, this result implies that the perturbation is irrelevant
at the decoupled XY FP.  However, since $\omega_u\equiv -y_u$ is very
small, the scaling corrections, that behave as $\xi^{-\omega_u} =
\xi^{-0.0225}$, decay very slowly.

If the coupling $w$ is negative, only one bosonic component is
expected to condense.  In this case the transition is characterized by
the symmetry breaking (\ref{symmp2}).  Therefore, if the BEC
transition is continuous, the critical behavior must belong to another
universality class, different from the XY one. The corresponding FP
has been extensively studied within the equivalent O(2)$\otimes$O(2)
LGW theory.\cite{Kawamura-88,PRV-01,CPPV-04} Estimates of the
corresponding critical exponents are: (i) $\nu=0.57(3)$ and
$\eta=0.09(1)$ from the resummation of the six-loop expansion within
the massive zero-momentum scheme;\cite{PRV-01} (ii) $\nu=0.65(6)$ and
$\eta=0.09(4)$ from five-loop calculations within the minimal
subtraction renormalization scheme.\cite{CPPV-04} These theoretical
results are also supported by experiments, see e.g.,
Ref.~\onlinecite{PV-02} and references therein.\cite{foot-controversy}
Therefore, the stable FP of the LGW theory (\ref{phi4s}) with
attraction domain in the region $w<0$ is characterized by the critical
exponents $\nu\approx 0.6$ and $\eta\approx 0.1$.  Of course, models
which are outside the attraction domain of the FP are expected to
undergo a first-order phase transition.

We finally mention that the critical exponents of the unstable
O(4)-symmetric FP along the separatrix $w=0$ are
\cite{HV-11,Hasenbusch-01} $\nu=0.750(2)$ and $\eta=0.0360(3)$.  This
FP is unstable because the spin-4 perturbation present when $w\not=0$
has positive RG dimension $y_w=0.125(5)$ at the O(4)
FP.~\cite{HV-11,CPV-03} Thus, an O(4) critical behavior can only be
observed by performing a proper tuning of the parameters of the model.

In the following sections we study model (\ref{sBH}) in the hard-core
$V\to +\infty$ limit.  Since the intra-species $V$ is naively related
to the quartic coupling $g$, we expect $g$ to be large in this limit,
so that $w\equiv g-u>0$.  Therefore, the BEC transition should be
characterized by the simultaneous condensation of both components, and
controlled by the decoupled FP, with a low-temperature phase in which
both components condense.  Scaling corrections, due to the on-site
density-density interaction between the two components, decay very
slowly.  We also expect such corrections to be larger when the
inter-species on-site interaction is attractive, i.e., for $U<0$,
while they should be small in the opposite case $U>0$. This is also
suggested by the fact that the hard-core 2BH model
becomes equivalent to the one-component model in the limit $U\to
+\infty$.  Therefore, for $U\to +\infty$ we expect a standard XY
transition without slowly-decaying $O(\xi^{-\ou})$ scaling
corrections.

\subsection{Multicritical behavior for two unequal bosonic gases}
\label{ungas}

\begin{figure}
\vskip5mm
\includegraphics[width=4.2cm]{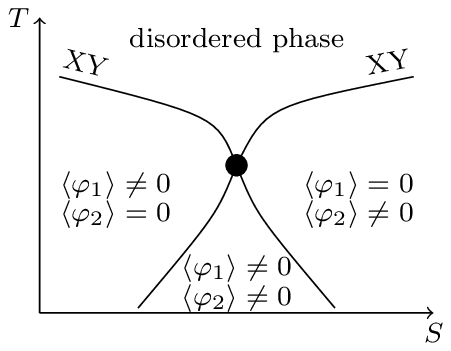}
\includegraphics[width=4.2cm]{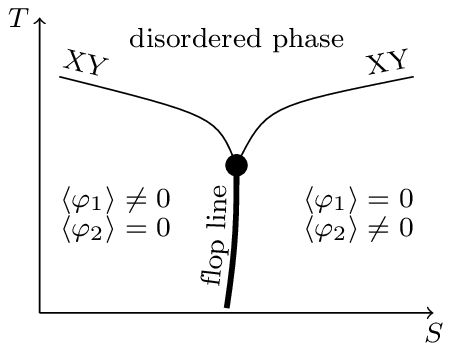} 
\caption{Different phase diagrams for two interacting
  bosonic gases.  Thin lines indicate
  continuous transitions, while the thick line represents first-order
  transitions.  RG analyses predict that only in the tetracritical case
  (left panel) may one have a continuous multicritical behavior at the 
  intersection of the transition lines.  }
\label{mcpd}
\end{figure}

We now discuss a system of two unequal bosonic species, such as that
described by the more general BH model
\begin{eqnarray}
H &=& -  \sum_{s} t_s \sum_{\langle {\bm xy}\rangle} 
(b_{s\bm{x}}^\dagger b_{s\bm{y}}^{} + {\rm h.c}) 
-  \sum_{s} \mu_s \sum_{\bm{x}} n_{s\bm{x}} 
\quad\label{sBHg}\\
&+&  \sum_{s} \frac{1}{2}
 V_{s} \sum_{\bm{x}} n_{s\bm{x}} (n_{s\bm{x}}-1) 
+ U \sum_{\bm{x}} n_{1\bm{x}} n_{2\bm{x}}. 
\nonumber
\end{eqnarray}
In this case we expect a more complex phase diagram, showing various
phases with transition lines along which only one bosonic component
condenses, and multicritical points (MCPs), where the critical
behavior arises from the competition of the two distinct U(1)
orderings.  More specifically, a MCP should be observed at the
intersection of the normal-to-superfluid transition lines where one of
the components condenses.

The LGW theory describing the competition of the two different U(1)
orderings of the model (\ref{sBHg}) is obtained by constructing the
most general $\Phi^4$ theory of two complex fields $\varphi_{s}({\bm
  x})$, with an independent U(1) symmetry for each component, without
exchange symmetry.  It reads
\begin{eqnarray}
{\cal H}_{\rm LGW} &=&\int d^3 x\,\Bigl[
\sum_{s,\mu} |\partial_\mu \varphi_s |^2  +  
\sum_s r_s |\varphi_s|^2 \label{mcphi4s}\\
&+& \sum_s g_s |\varphi_s|^4 
+ 2u\, |\varphi_1|^2 |\varphi_2|^2\Bigr],
\nonumber
\end{eqnarray}
where now we have two quadratic parameters $r_1$ and $r_2$ and three
quartic parameters $g_1$, $g_2$, and $u$.  The multicritical behavior
arising from the competition of the two distinct U(1) orderings is
determined by the RG flow when both quadratic parameters $r_1$ and
$r_2$ are simultaneously tuned to their critical values, keeping the
quartic parameters $g_1$, $g_2$ and $u$ fixed.

The phase diagram of the most general theory, in which the associated
symmetries are O($n_1$) and O($n_2$), has already been investigated
within the mean-field approximation.\cite{NKF-74,KNF-76,CPV-03}
Several different phase diagrams have been identified, characterized
by three or four transition lines meeting at a MCP, characterized by
the presence or the absence of a mixed phase, in which both fields
condense.  In Fig.~\ref{mcpd} we show the phase diagrams corresponding
to the case of two coupled U(1)-symmetric theories, in the $T$-$S$
plane where $S$ represents a second relevant parameter (for instance,
the difference of the chemical potentials of the two species) that
must be tuned to obtain the multicritical behavior.  In the LGW theory
the two behaviors are determined by the sign of $\Delta\equiv g_1 g_2
- u^2$. If $\Delta>0$, four critical lines meet at the MCP
(tetracritical behavior), as in the left panel of Fig.~\ref{mcpd},
while, if $\Delta<0$, two critical lines and one first-order line
(bicritical behavior) are present, see the right panel of
Fig.~\ref{mcpd}.

The sign of $\Delta$ also controls the nature of the behavior at the MCP.
The FT analysis~\cite{Aharony-02,CPV-03} of the LGW theory
(\ref{mcphi4s}) shows that the system undergoes a first-order transition 
at the MCP for $\Delta < 0$, i.e., in the bicritical case, as no stable 
FP is found in this parameter region. In the opposite case (tetracritical 
phase diagram) a continuous transition is possible at the MCP, controlled
by the decoupled XY FP.

\section{Mean-field phase diagram of the 2BH model}
\label{mfan}

The phase diagram of the 2BH model (\ref{sBH}) can be studied
in the
mean-field approximation, using
\begin{eqnarray}
  b^\dag_{s\bm{x}} b_{s\bm{y}}
  &&= \left[ (b^\dag_{s\bm{x}} - \phi_s^*) + \phi_s^* \right] 
     \left[ (b_{s\bm{y}} - \phi_s) + \phi_s \right] \nonumber \\ 
  && \approx \phi_s b^\dag_{s\bm{x}} + \phi_s^* b_{s{\bm y}} - |\phi_s|^2,
\label{approx-meanfield}
\end{eqnarray}
where $\phi_{s}=\langle b_{s \bm x }\rangle$ are two complex
space-independent variables, that play the role of order parameters at
the BEC transitions.  Approximation (\ref{approx-meanfield}) allows us
to rewrite Hamiltonian (\ref{sBH}) as a sum of decoupled one-site
Hamiltonians
\begin{eqnarray} 
 && H[\phi_s] = 
     - 2DJ \sum_{s} \left(\phi_s 
         b^\dag_{s} + \phi_s^* b^{}_{s} 
         - |\phi_s|^2 \right) \label{eq:mf_hamiltonian}  \\ 
    && - \mu \sum_{s} n_{s}  
      + \frac{V}{2} \sum_s 
         n_{s} (n_s - 1) 
+ U n_{1}n_{2}.    \nonumber
\end{eqnarray} 
The spectrum of the theory is invariant under the redefinition $b_s
\to U_s b_s$, where $U_s$ are two arbitrary phases. Therefore, there
is no loss of generality if the two parameters $\phi_s$ are assumed to
be real.  They are determined by minimizing the single-site free
energy with respect to $\phi_s$.

In the hard-core limit, since $n_{s}=0,\,1$, the mean-field
Hamiltonian (\ref{eq:mf_hamiltonian}) is defined on a Hilbert space of
dimension 4.  In the soft-core case, we may have any occupation
number, so that the Hilbert space has infinite dimension. In practice,
we only consider states such that $n_s \le n_{\rm max}$, checking that
typical occupation numbers are significanly lower than $n_{\rm max}$
and verifying that the results are stable with respect to changes of
the cut-off $n_{\rm max}$.

\begin{figure}
 \includegraphics{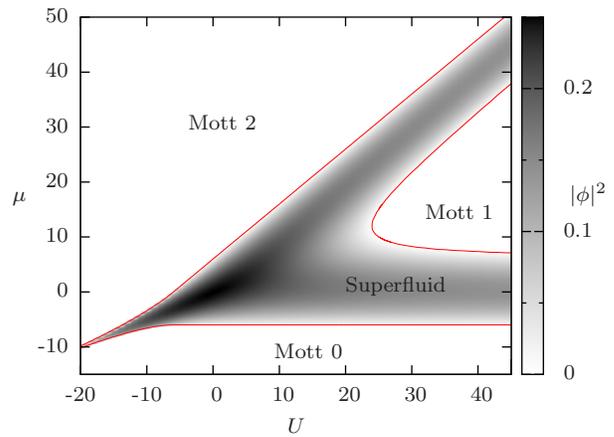}
 \caption{ Zero-temperature $U$-$\mu$ phase diagram of the 
2BH model in the hard-core $V\to\infty$ limit. 
   }
\label{fig:mf_hardcore_heatmap}
\end{figure}

\begin{figure}
 \includegraphics{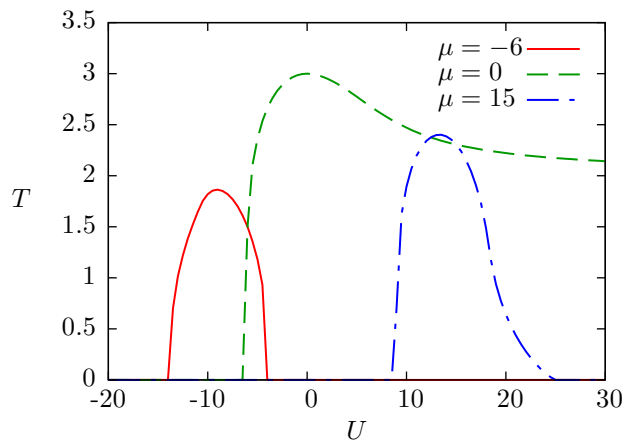}
 \caption{Phase diagram of the hard-core 2BH model for
   $\mu=-6,\,0,\,15$ in the $U$-$T$ plane. In particular, we show
   the normal-to-superfluid transition lines.  }
\label{fig:mf_hardcore_ft}
\end{figure}

\begin{figure}
 \includegraphics{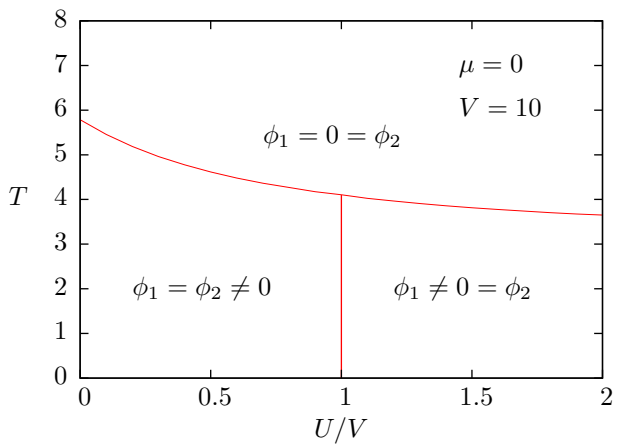}
 \caption{Phase diagram of the 2BH model for $V=10$ and $\mu=0$ in the
   $U$-$T$ plane.  The transitions along the line $U/V=1$ separating
   the two low-temperature phases are of first order. }
\label{fig:mf_softcore_ft}
\end{figure}

In Fig.~\ref{fig:mf_hardcore_heatmap} we report $|\phi_1|^2 =
|\phi_2|^2$ for the hard-core model at $T=0$ as a function of $U$ and
$\mu$. The parameter space is divided into four regions: a central one
(superfluid domain) in which $|\phi|^2\not = 0$, and three regions in
which $|\phi|^2 = 0$ that differ by the value of the total occupation
number $n = n_1 + n_2$. There is a vacuum region in which $n = 0$ and
two Mott incompressible phases with $n=1$ and $n=2$.  The $n=1$ Mott
phase presents several interesting features related to the {\em
  isospin} degrees of freedom per site, which may be described by
effective low-energy spin
Hamiltonians.\cite{KS-03,DDL-03,AHDL-03,ICSG-05} The superfluid domain
for $\mu<0$, separating the $n=0$ and $n=2$ Mott phases, extends in a
strip around the line $U=2\mu$ that gets narrower as $\mu$ decreases
(the $n=2$ Mott phase cannot include the line $U=2\mu$ where the
on-site interactions would cancel for $n_{1}=n_{2}=1$). This
mean-field phase diagram appears quite similar to that of the 1D
Hubbard model which can be exactly solved, see, e.g.,
Refs.~\onlinecite{1DHM,ACV-14} (we recall that the 1D hard-core 2BH
model can be exactly mapped onto the 1D fermion Hubbard model).

The mean-field calculations can be extended to finite temperatures by
minimizing the free-energy density
\begin{equation}
 f(\phi, \beta) = -\frac{1}{\beta} \ln \mathcal{Z}(\phi, \beta), \quad 
 \mathcal{Z}(\phi, \beta) = \sum_i e^{-\beta E_i},
\end{equation}
with respect to the variational parameters $\phi_s$.  $E_i$ are the
energy levels of the single-site Hamiltonian.  The finite-temperature
phase boundaries are obtained by looking for the smallest value of
$\beta\equiv 1/T$ for which, at any given $U$ and $\mu$, $\phi_{s}$
assume non-zero values.

The phase diagrams in the hard-core limit are shown in
Fig.~\ref{fig:mf_hardcore_ft} for some values of the chemical
potential $\mu$.  In all cases the low-temperature superfluid phases
are characterized by the simultaneous condensation of both components.
In particular, for $\mu=0$, the case we will investigate numerically,
there is a low-temperature superfluid phase for $U>-6$, while the
system is always in the normal phase for $U<-6$.

In Fig.~\ref{fig:mf_softcore_ft} we show the phase diagram for a
finite value of the intra-species coupling, $V=10$, and for
$\mu=0$. In this case we have two different low-temperature phases:
when $V>U$ both components condense as in the hard-core limit, while
for $V<U$ only one component condenses, breaking the exchange
symmetry.  These different condensed phases are separated by a
first-order transition line at $U=V$.  These results are completely
consistent with the predictions obtained by analyzing the
corresponding LGW theory (\ref{phi4s}), see Sec.~\ref{ftrgantwof}.

\section{Numerical study of two-component hard-core  bosons}
\label{qmc}

We now check some of the theoretical predictions of the previous
sections. We present a numerical analysis of the critical behavior of
hard-core 2BH model (\ref{sBH}). As discussed in
Sec.~\ref{ftrgantwof}, we expect a critical transition in the 3D XY
universality class with a simultaneous condensation of both
components. Correspondingly, we have \cite{CHPV-06} $\nu=0.6717(1)$,
$\eta = 0.0381(2)$.  However, the asymptotic behavior is approached
with slowly-decaying scaling corrections, that behave as
$\xi^{-\omega_u}$ with $\omega_u=0.0225(4)$.  These corrections are
expected to give rise to significant effects when the inter-species
on-site interaction is attractive, i.e., for $U<0$, while they may be
negligible in the repulsive case.  In the following we provide
numerical evidence for this scenario.

\subsection{Monte Carlo simulations and observables}
\label{mcobs}

We perform quantum Monte Carlo (QMC) simulations of the
hard-core 2BH model at zero chemical potential $\mu=0$, on cubic $L^3$
lattices with periodic boundary conditions, for $L$ up to 64.  We use
the directed operator-loop algorithm,\cite{SK-91,SS-02,DT-01} which is
a particular algorithm using the stochastic series expansion (SSE)
method.~\cite{footnote-mc} In the simulations we determine the
helicity modulus and the second-moment correlation length.  The
helicity modulus $\Upsilon$ is the response of the system to a twist
of the boundary conditions.  It can be obtained from the linear
winding number $w_i$ along the $i^{\rm th}$ direction,
\begin{equation}
\Upsilon =  \frac{\langle w_i^2 \rangle}{L},
\qquad w_i = \frac{N_i^+ - N_i^-}{L},
\label{ulw}
\end{equation}
where $N_i^+$ and $N_i^-$ are the number of non-diagonal operators
which move the particles respectively in the positive and negative
$i^{\rm th}$ direction.  The second-moment correlation length $\xi$
can be conveniently defined from the lattice Fourier transform
${\widetilde{G}(\bm{p})}$ of the two-point correlation function
$G({\bm x}-{\bm y})=\langle b^{\dagger}_{\bm{x}} \, b_{\bm{y}}\rangle$, as
\begin{equation} \label{eq:xi_definition} 
 \xi^2 \equiv \frac{1}{4 \sin^2 (\pi/L)} 
      \frac{\widetilde{G}({\bm 0}) - \widetilde{G}({\bm p})} 
      {\widetilde{G}({\bm p})}, 
\end{equation} 
where ${\bm p}=(2\pi/L,0,0)$.

To determine the critical behavior, we perform a finite-size scaling
(FSS) analysis of the RG invariant quantities $R_\Upsilon = \Upsilon
L$ and $R_\xi = \xi/L$ (we generically denote them as $R$).  Close to
the transition point $\beta\equiv 1/T=\beta_c$, they behave as
\begin{equation}
\label{eq:Rscaling} 
 R(\beta,L) = f_R(\tau L^{1/\nu}) + 
    L^{-\omega_1} g_R(\tau L^{1/\nu}) +
O(L^{-\omega_2},L^{-2\omega_1}),
\end{equation} 
where $f_R(x)$ is universal apart from a rescaling of its argument,
$\tau\equiv 1- \beta/\beta_c$, $\nu$ is the correlation-length
exponent, $\omega_i$ ($0<\omega_1<\omega_2<...$) are the exponents
controlling the scaling corrections to the asymptotic behavior, which
are associated with the irrelevant perturbations at the stable FP.

The scaling equation (\ref{eq:Rscaling}) implies that data for
different values of $L$, in particular $L_1=L$ and $L_2=2L$, cross at
a given $\beta_{\rm cr}(R;L_1,L_2)$, which approaches $\beta_c$ for
$L\to\infty$. More precisely,
\begin{equation}
\beta_{\rm cr}(R;L,2L) = \beta_c + O(L^{-1/\nu-\omega_1}). \label{betaL}
\end{equation}
Moreover, the value of $R$ for $\beta = \beta_{\rm cr}$ approaches the
universal critical value $R^*=f_R(0)$, i.e.
\begin{equation}
R(\beta_{\rm cr},L) = R^* + \sum_{n=1} b_{1n} L^{-n\omega_1} +
\sum_{n=1} b_{2n} L^{-n\omega_2} + \ldots
\label{rlasy}
\end{equation}

\subsection{QMC results for $U=0$}
\label{u0res}

\begin{figure} 
\includegraphics{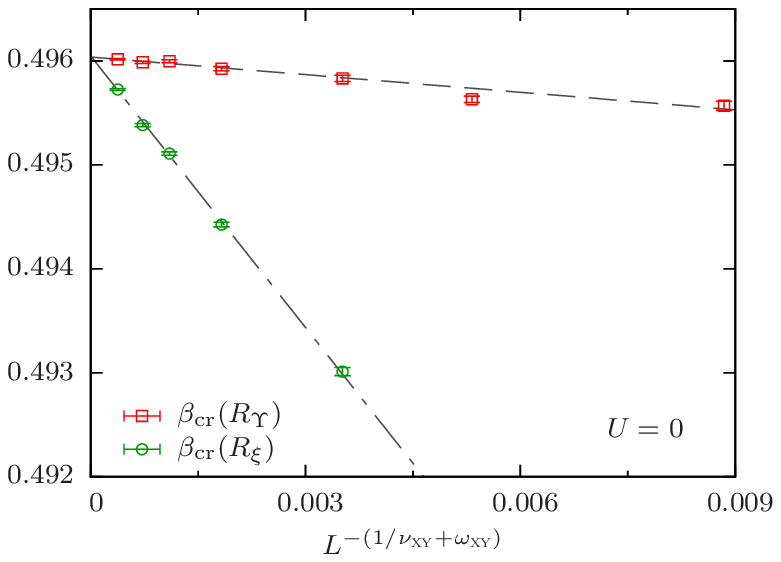} 
\includegraphics{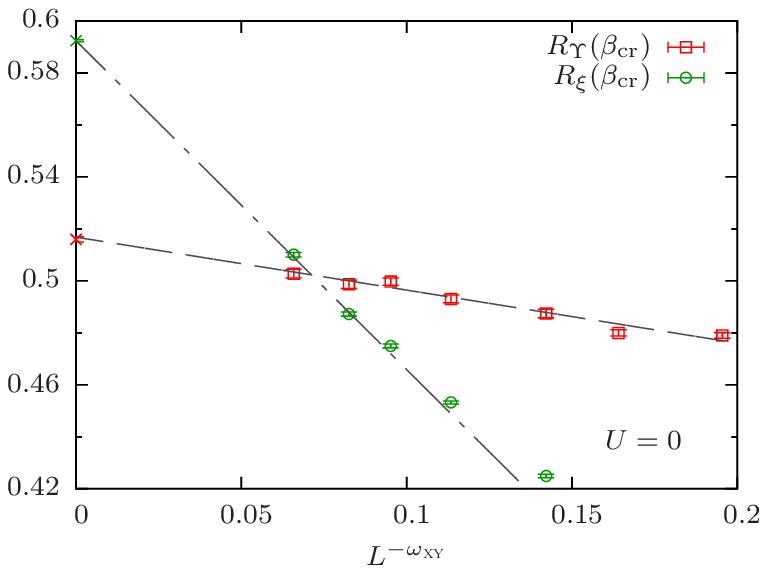}
\caption{ (Top) Crossing points $\beta_{\rm cr}(R_\Upsilon;L,2L)$ and
  $\beta_{\rm cr}(R_\xi;L,2L)$ for $U=0$ and $\mu=0$. They are plotted
  versus $L^{-1/\nuxy-\oxy} = L^{-2.27}$, which is the expected
  behavior of the leading scaling corrections.  The dashed lines
  correspond to linear fits of the data for the largest available
  lattices.  (Bottom) Estimates of $R_\Upsilon$ and $R_\xi$ at the
  crossing points, versus $L^{-\oxy} = L^{-0.785}$.  The dashed lines
  correspond to linear fits to $R^* + c L^{-\oxy}$ with $R^*$ fixed to
  its 3D XY value [$R_\Upsilon^* = 0.516(1)$ and $R_\xi^* =
    0.5924(4)$], which is reported along the ordinate axis (crosses).}
\label{fig:fit_U0}
\end{figure}

To begin with, we consider the hard-core 2BH model for $U=0$,
representing two decoupled and identical single-component hard-core BH
models.  The results will then be compared with those obtained for
$U\neq 0$.

In this case we have a robust theoretical prediction for its critical
behavior at the BEC transition: it belongs to the 3D XY universality
class, described by a standard U(1)-symmetric $\Phi^4$ theory with one
complex order parameter.  \cite{CN-14,CTV-13,PV-02} The leading
scaling corrections decay with exponent $\omega_1 = \oxy = 0.785(20)$,
and the asymptotic critical values of $R_\Upsilon$ and $R_\xi$ are
$R_\Upsilon^* = 0.516(1)$ and $R_\xi^* = 0.5924(4)$,
respectively.~\cite{CHPV-06} Numerical evidence of this critical
behavior has already been reported in Refs.~\onlinecite{CTV-13,CN-14}.

We determine the crossing points~\cite{footnote_cr} $\beta_{\rm
  cr}(R;L,2L)$ for $L$ up to 32.  Results are shown in the top panel
of Fig.~\ref{fig:fit_U0}.  The behavior for $L\to\infty$ is consistent
with the expected $O(L^{-1/\nuxy - \oxy})$ scaling corrections.
Linear fits of the data for $L\gtrsim 20$ give $\beta_c =
0.496035(10)$,~\cite{footnote_betac} which is in agreement with, and
slightly improves, earlier estimates.\cite{CN-14} The values of
$R_\Upsilon$ and $R_\xi$ at the crossing points are reported in the
lower panel of Fig.~\ref{fig:fit_U0}.  They show the expected
asymptotic behavior $R(\beta_{\rm cr},L) = R^* + c
L^{-\oxy}$. Moreover, the extrapolated values are consistent with the
best available estimates for the 3D XY universality class. For
example, linear fits of the data for $L\gtrsim 20$ give $R_\Upsilon^*
= 0.516(4)$, which is in agreement with the best available
estimate~\cite{CHPV-06} $R_\Upsilon^*=0.516(1)$ of the 3D XY
universality class.

\subsection{QMC results for $U\not=0$}
\label{otherU}

We now present results for the hard-core 2BH model for
$U\not=0$. In this case we should consider the slowly-decaying scaling
corrections of order $L^{-\omega_u}$ predicted in
Sec.~\ref{ftrgantwof}, which are expected to give rise to 
significant systematic
deviations, at least for negative $U$. Since $\omega_u = -0.0225$,
these deviations are hardly detectable numerically. Indeed, in our
range of values of $L$, $8\le L\le 64$, $L^{-\omega_u}$ varies only by
5\%, hence it is very difficult to distinguish it from a constant
term. In practice, unless data are extremely precise, any FSS analysis
is unable to determine the leading scaling function $f_R(\tau
L^{1/\nu})$ appearing in Eq.~(\ref{eq:Rscaling}). The extrapolation of
the data to $L\to \infty$ would identify the asymptotic behavior with
that given by
\begin{equation}
\widetilde{f}_R (\tau L^{1/\nu}) = f_R (\tau L^{1/\nu}) + 
A \, g_R(\tau L^{1/\nu}),
\label{fakesca}
\end{equation}
where the slowing-decaying factor $L^{-\omega_u}$ is effectively
replaced with some kind of average $A\equiv [L^{-\omega_u}]_{\rm av}$
in the considered range of values of $L$.  This observation suggests
that the analysis based on the large-$L$ extrapolation of the crossing
points should be able to determine correctly $\beta_c$ and $\nu$. On
the other hand, the extrapolation of the data for $R_\Upsilon $ and
$R_\xi$ at the critical point would give $\widetilde{f}_R (0)$, which
differs from the correct asymptotic estimate. In other words, we cannot
rely on the values of the RG invariant quantities at the critical
point to identify the universality class. In this discussion we have
assumed that there is only a single slowly-decaying correction term,
i.e., $L^{-\omega_u}$, which is, however, not the case. RG also
predicts the presence of correction terms proportional to $L^{-n
  \omega_u}$ for any integer $n$, cf. Eq.~(\ref{rlasy}), which makes
the analysis of the corrections even more difficult. Finally, note
that corrections of order $L^{-\oxy}$, the leading ones in the
single-species model, are also expected.

\begin{figure} 
\includegraphics{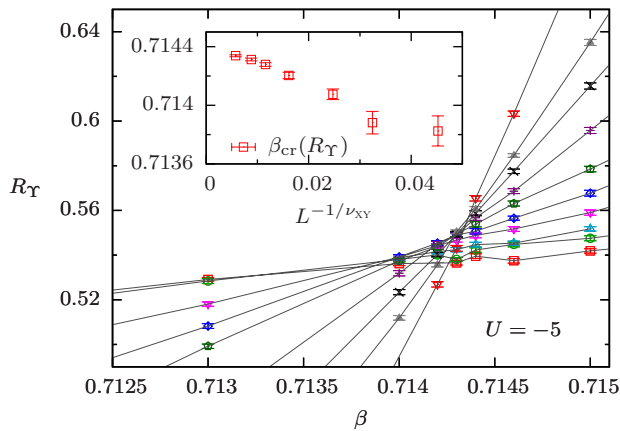}
\caption{QMC estimates of $R_\Upsilon$ for $U=-5$ and $\mu=0$.  The
  inset shows the crossing points $\beta_{\rm cr}(R_\Upsilon;L,2L)$
  versus $L^{-1/\nuxy}$.  }
\label{fig:rawY_U-5}
\end{figure}

\begin{figure}
\includegraphics{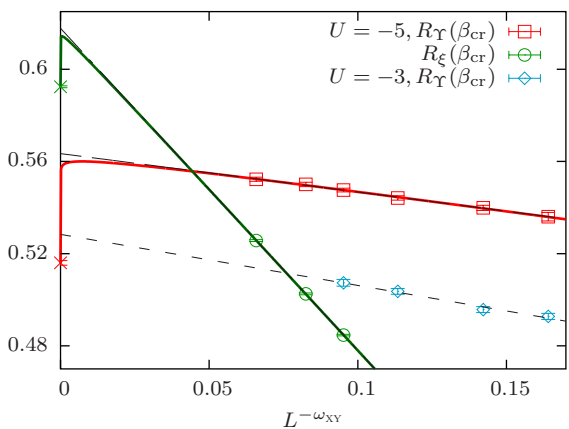}
\caption{$R_\Upsilon$ and $R_\xi$ at the crossing points obtained
  using data for lattices of sizes $L$ and $2L$.  Here $\mu=0$, $U=-5$
  and $U=-3$.  The straight lines correspond to linear fits to
  $R_\#(\beta_{\rm cr},L)=R_\#^* + c L^{-\oxy}$ with $R_\#^*$ and $c$
  as free parameters (for $U=-5$ we obtain $R_\Upsilon(\beta_{\rm
    cr},L) = 0.563 - 0.17 L^{-\oxy}$ for $L\ge 12$ and
  $R_\xi(\beta_{\rm cr},L) = 0.618 -1.40L^{-\oxy}$ for $L\ge 20$). We
  also report the results of the fits to $R_\#(\beta_{\rm
    cr},L)=R_\#^* + c_u L^{-\omega_u} + c L^{-\oxy}$, in which
  $R_\#^*$ is fixed to the 3D XY critical value and $c_u$ and $c$ are
  free parameters (we obtain $R_\Upsilon(\beta_{\rm cr},L) = 0.516 +
  0.052 L^{-\ou} - 0.18 L^{-\oxy}$ and $R_\xi(\beta_{\rm cr},L) =
  0.5924 + 0.028 L^{-\ou} -1.41 L^{-\oxy}$ for $U=-5$).  The crosses
  on the vertical axis mark the best estimates of $R^*$ for the 3D
  universality class.}
\label{fig:R_U-5}
\end{figure}

We first consider the attractive hard-core model with $U=-5$ and
$\mu=0$.  Fig.~\ref{fig:rawY_U-5} reports the estimates of
$R_\Upsilon$ as a function of $\beta$, which clearly show crossing
points between $\beta=0.714$ and $\beta=0.715$.  The crossing points
$\beta_{\rm cr}(R_\Upsilon;L,2L)$ up to $L=32$ are shown in the
inset. We stress that their determination requires no prior knowledge
on the nature of the transition, and it is therefore completely
unbiased.  They clearly appear to converge to a critical value
$\beta_c$.  The precision of the data does not allow us to distinguish
the expected $O(L^{-1/\nuxy-\omega_u})$ approach to the asymptotic
value, as predicted by theory, from the $O(L^{-1/\nuxy-\oxy})$
approach at $U=0$.  Nevertheless, we obtain a reasonably precise
estimate $\beta_c=0.7144(1)$, where the error also accommodates the
difference of the extrapolations using the two ansatzes.  Analogous
results, although less precise, are obtained from the $R_\xi$ data.

The values of $R_\Upsilon$ and $R_\xi$ at the crossing points are
shown in Fig.~\ref{fig:R_U-5}.  They show an apparently linear
behavior when plotted against $L^{-\oxy}$, as in the $U=0$
case. However, an extrapolation using the ansatze $R^* + c L^{-\oxy}$,
which appears consistent with the data, gives critical values for
$R_\Upsilon^*$ and $R_\xi^*$ which are definitely different from those
of the 3D XY universality class. For example, we obtain
$R_\Upsilon^*=0.563(2)$ and $R_\xi^*=0.625(5)$ with an acceptable
$\chi^2/{\rm DOF}$ (DOF is the number of degrees of freedom of the
fit), which differ significantly from the XY estimates
$R_\Upsilon^*=0.516(1)$ and $R_\xi^*=0.5924(4)$.  In view of the
previous discussion, this discrepancy should be expected, because of
the presence of the slowly-decaying $O(L^{-\ou})$ corrections with
$\ou=0.0225(4)$. For instance, the data in Fig.~\ref{fig:R_U-5} can
also be nicely fitted to
\begin{equation}
R(\beta_{\rm cr}) = R^* + aL^{-\ou} + bL^{-\oxy},
\label{fitR}
\end{equation}
with $R^*$ fixed to its XY value, as shown in
Fig.~\ref{fig:R_U-5}. Note that fits which include the $O(L^{-\ou})$
corrections are hardly distinguishable from the ones that assume the
leading correction to be $L^{-\oxy}$ in the range of $L$ for which the
MC data are available. This confirms that disentangling the correction
term from the leading constant is extremely hard, requiring accurate
computations for very large lattice sizes.  It should be stressed that
the fit to the ansatze (\ref{fitR}) is only an exercise. It is
presented to make plausible that the transition is in the XY
universality class, even though a naive fit of the data provides a
different value for $R^*$.  Indeed, fit (\ref{fitR}) is not
conceptually correct unless $L$ is enormously large. RG predicts
corrections of order $L^{-n\ou}$ for any integer $n$, that are as
relevant as the leading one in our range of values of $L$.

To further check the above-reported results, we repeat the FSS
analysis at fixed $\beta=0.7144$, varying the on-site coupling $U$
which now takes the role that $\beta$ had in the previous analysis. An
analogous FSS analysis of data up to $L=48$ gives $U_c=-4.9999(3)$,
which is perfectly consistent with the FSS analysis at fixed $U$.
Moreover, the values of $R_\Upsilon$ and $R_\xi$ at the crossing
points in the variable $U$ are hardly distinguishable from those
appearing in Fig.~\ref{fig:R_U-5} at the corresponding values of $L$.

We have also performed a FSS analysis of data for $U=-3$, up to
$L=40$, obtaining $\beta_c=0.5390(2)$.  The data of $R_\Upsilon$ at
the crossing points $\beta_{\rm cr}(R_\Upsilon;L,2L)$ are also shown
in Fig.~\ref{fig:R_U-5}. As for $U=-5$, they appear to behave linearly
with respect to $L^{-\oxy}$, and again they extrapolate to a value
that is significantly larger than $R_\Upsilon^*\approx 0.516$.  Such a
deviation is smaller than that obtained for $U=-5$, confirming that
the discrepancies cannot be interpreted as due to the presence of a
new universality class. In that case, indeed, one would obtain the
same extrapolated value for both values of $U$. Instead, the results
are consistent with our RG predictions: the discrepancies increase
with $|U|$, which is exactly what should be expected if they are
related to the slowly-decaying corrections due to the interspecies
interaction.

We finally mention that, as already anticipated in
Sec.~\ref{ftrgantwof}, the scaling corrections induced by the
density-density on-site interaction turn out to be small when the
interspecies interaction is repulsive, that is for $U>0$. We have
considered two values of $U$, $U=1$ and 10, and lattices up to
$L=40$. In both cases, we obtain results for $R^*_\Upsilon$ and
$R^*_\xi$ that are in agreement with the XY values. Apparently, the
slowly-decaying corrections are negligible, being at most of the size
of the statistical errors. Note that these corrections vanish for
$U=0$ (the two models are decoupled) and also for $U\to +\infty$ (the
model is equivalent to the hard-core BH model for a single boson,
hence it has a standard XY transition without the $L^{-\ou}$ scaling
corrections).  Apparently, they keep on being small for all
intermediate values of $U$.

\subsection{Phase diagram of the hard-core 2BH model at $\mu=0$}
\label{phdia}

\begin{figure} 
 \centering \includegraphics{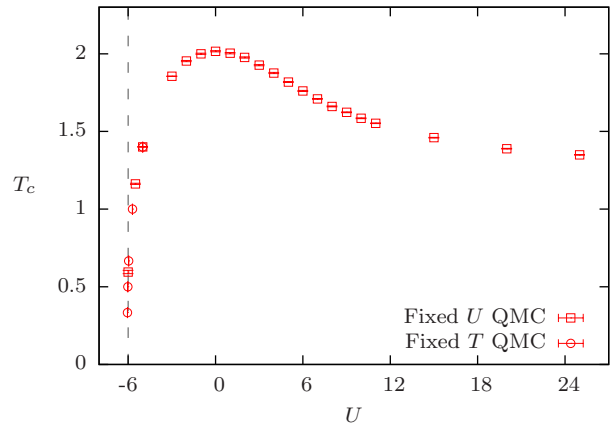}
 \caption{Transition line of the hard-core 2BH model for
   $\mu=0$, as obtained by the FSS analysis of QMC data.  We plot
   $T_c\equiv 1/\beta_c$ versus $U$. Most estimates are obtained in
   simulations at fixed $U$ in which $T$ is varied.  Close to $U=-6$,
   where the transition line is almost parallel to the $T$ axis, we
   performed simulations at fixed temperature, determining the
   critical coupling $U_c$.}
 \label{fig:phdia}
\end{figure}

We determine the $U$ dependence of the normal-to-superfluid transition
line by repeating the FSS analysis for other values of $U$.  This is
done with less accuracy, using data up to $L=24$.  The results are
shown in Fig.~\ref{fig:phdia}.  The phase diagram is quite similar to
that obtained in the mean-field approximation. In particular, the
low-temperature superfluid phase disappears for $U\lesssim -6$, very
close to the mean-field result $U= -6$.  Indeed, as shown by the
zero-temperature mean-field phase diagram shown in
Figs.~\ref{fig:mf_hardcore_heatmap} and \ref{fig:mf_hardcore_ft},
$U=-6$ is the location of the quantum transition between the
superfluid and $n=2$ Mott phase.

\begin{figure} 
 \centering 
 \includegraphics{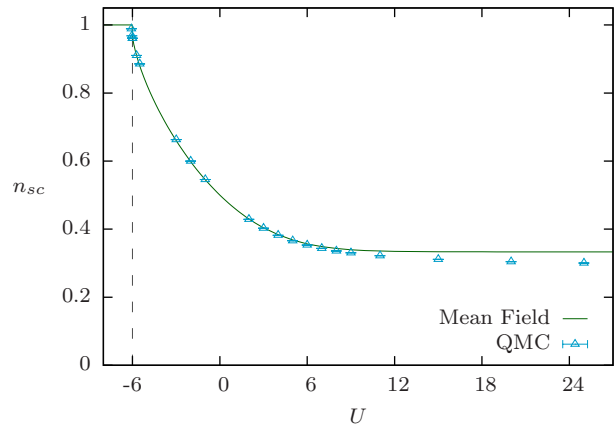} 
 \caption{The single-species particle density $n_s\equiv \langle
   n_{s{\bm x}} \rangle$ along the critical line, as obtained by QMC
   and mean-field calculations. }
 \label{fig:density}
\end{figure}

Finally, we discuss the behavior of the particle densities $n_s \equiv
\langle n_{s\bm{x}} \rangle $ at the BEC transition.  Their leading
behavior at the BEC transitions arises from the analytical background
terms, while the universal power laws related to the critical behavior
are subleading. Indeed, standard RG arguments predict the asymptotic
behavior
\begin{equation}
n_s  \approx   f_a(\tau)  + L^{-y_n} f_s(\tau L^{1/\nu})  
\label{rhofss}
\end{equation}
when varying the reduced temperature $\tau=1-\beta/\beta_c$ 
keeping fixed the model parameters.
In Eq.~(\ref{rhofss}),
$f_a$ is a nonuniversal analytic function; $y_n$
is the RG dimension of the particle density operator $n_{s\bm{x}}$,
which is  given by $y_n=3-1/\nu=1.5112(2)$ at the decoupled XY FP; 
  $f_s$ is a universal function apart from a factor and a rescaling of its
  argument. Therefore, the critical densities at $T_c$
  are expected to approach a nonuniversal constant in the
  large-$L$ limit.  In Fig.~\ref{fig:density} we show 
 the    large-$L$ extrapolations of the particle-density data 
 at $T_c$ versus the on-site coupling $U$.
  The comparison with the mean-field
  computations of Sec.~\ref{mfan} shows that the mean-field
  approximation of the particle density is quite accurate.

\section{Conclusions}
\label{conclu}

We investigate BEC in 3D two-component bosonic systems.  In particular
we consider two interacting identical bosonic gases, described by the
2BH model (\ref{sBH}), which may be interpreted as a lattice
two-component bosonic system.  We study the phase diagram and the
critical behavior by RG, mean-field, and numerical methods.

Our RG analysis is based on a LGW theory with two complex scalar
fields (associated with the two bosonic components), which has the
same symmetry as that of the bosonic system.  In the case of two
identical components with density-density interactions, the relevant
global symmetry is ${\mathbb Z}_{2,e} \otimes {\rm U}(1) \otimes {\rm
  U}(1)$.  The mean-field analysis predicts two different types of
low-temperature phases.  Depending on the values of the on-site
inter-species $U$ and intra-species $V$ couplings, one may have (i) a
phase in which the exchange symmetry is conserved and both components
condense or (ii) a phase in which only one component condenses, thus
breaking the ${\mathbb Z}_{2,e}$ exchange symmetry.

In case (i), which is generically expected for $V\gtrsim U$, the
transition belongs to the 3D XY universality class. More precisely,
the critical behavior is controlled by a decoupled XY FP, implying an
asymptotic decoupling of the critical modes associated with the
bosonic components.  The density-density interaction between these two
components turns out to be an irrelevant perturbation at this FP.  It
does not affect the asymptotic behavior but gives rise to
slowly-decaying scaling corrections, that behave as $\xi^{-\omega_u}$,
where $\xi$ is the diverging length scale at transition and
$\omega_u=3-2/\nuxy=0.0225(4)$.  Of course, these slowly-decaying
effects are absent at the transition of a single bosonic
species,~\cite{CN-14,CTV-13,CR-12,CPS-07} where the leading scaling
corrections decay as $\xi^{-\oxy}$ with $\oxy\approx 0.78$.  The
presence of slowly decaying corrections makes an accurate check of the
asymptotic 3D XY critical behavior quite hard, essentially because one
needs to get very close to the critical point to make them negligible.
These predictions are supported by a FSS analysis of QMC data for the
2BH model (\ref{sBH}) in the hard-core limit, i.e., for $V\to\infty$.

The FT RG analysis predicts that the nature of the transition should
significantly change in the soft-core regime, i.e. when $V\lesssim
U$. In this case only one component is expected to condense.  The
corresponding symmetry breaking is therefore different, hence it leads
to a different universality class in the case of continuous
transitions.  We identify this universality class with that of the
chiral transition in frustrated two-component spin models with
noncollinear order,~\cite{Kawamura-88,PRV-01,CPPV-04} which has
$\nu\approx 0.6$ as correlation-length exponent.

Our RG study also predicts the possibility of a critical behavior with
an extended O(4) symmetry. However, such symmetry enlargement can only
be observed by tuning a further parameter beside the temperature.

It should be stressed that the different critical behaviors can only
be observed if the system is in the attraction domain of one of the
FPs. If this is not the case, the transition would be of first order.

We also extend our analysis to the more general case in which the two
bosonic components are not identical. The phase diagrams are expected
to be more complex, as shown in Fig.~\ref{mcpd}.  In particular, they
may or may not show a low-temperature mixed phase characterized by the
condensation of both components.  According to mean-field and RG
results, if the mixed phase is present, the phase diagram presents a
tetracritical point where four transition lines meet, and the
multicritical behavior is controlled by the decoupled XY FP.  When the
mixed phase is absent, thus the low-temperature phases are
characterized by the BEC of only one component, the transitions
between the two BEC phases must be first order.  In this case the
competition of the two U(1) orderings does not lead to a multicritical
behavior, since no stable FPs are found in the corresponding parameter
region; as a consequence the behavior at the intersection of the
transition lines is expected to show thermodynamic discontinuities
analogous to first-order transitions.

The RG analysis of BEC transitions in mixtures of bosonic gases can be
straightforwardly extended to the two-dimensional (2D) case.  In two
dimensions bosonic systems do not experience BEC. The low-temperature
phase of single species is characterized by a quasi-long range order,
where correlations decay algebraically at large distances, without the
emergence of a nonvanishing order parameter. The transitions to this
low-temperature phase are generally of the
Berezinskii-Kosterlitz-Thouless (BKT)
type,~\cite{Berezinski-72,KT-73,Kosterlitz-74} characterized by an
exponential increase of the correlation length.  The phase diagram of
mixtures of 2D bosonic systems may show BKT transitions such as those
of single bosonic species, and transitions related to the breaking of
the $\mathbb{Z}_{2}$ exchange symmetry.  A similar situation arises in
2D frustrated two-component spin models, see, e.g.,
Ref.~\onlinecite{HPV-05}, and references therein.  RG scaling
arguments analogous to those used for the 3D case allow us to infer
that the critical behavior of identical interacting hard-core bosonic
components is again controlled by a decoupled BKT FP.  Since the
energy operator is marginal at the BKT transition, i.e.  $y_e=0$, the
RG dimension of the density-density inter-species coupling is given by
$y_u = 2 y_e - d = -2$ ($d=2$ in this case).  Therefore, energy-energy
or density-density interactions between the bosonic species are
irrelevant also in two dimensions.  Unlike the 3D case, the
corresponding contributions get rapidly suppressed when approaching
the critical point, indeed they are $O(\xi^{-2})$. Therefore, we
expect that 2D identical hard-core bosonic components, such as the
hard-core 2BH model (\ref{sBH}) in two dimensions, undergo continuous
transitions characterized by decoupled BKT behaviors with
multiplicative and subleading logarithmic
corrections,\cite{AGG-80,PV-13} analogous to the case of a single
bosonic species~\cite{CNPV-13}.

We finally note that cold-atom experiments are usually performed in
inhomogeneous conditions, due to the presence of space-dependent
trapping forces which effectively confine the atomic gas within a
limited space region.~\cite{CW-02,Ketterle-02,BDZ-08} The trapping
potential is effectively coupled to the particle density. Thus, it can
be taken into account by adding a space-dependent trap term such as
\begin{eqnarray}
H_{\rm trap} = \sum_{\bm{x}} V_t({\bm x}) n_{s\bm{x}}, \label{bhmt}
\end{eqnarray}
to the BH Hamiltonian (\ref{sBH}), where $V_t$ is the space-dependent
potential associated with the external force.  For example, we may
consider $V_t(r)= (r/\ell_t)^2$, where $r\equiv |{\bm x}|$ is the
distance from the center ${\bm x}=0$ of the trap, which describes a
harmonic rotationally-invariant trap.  The inhomogeneity arising from
the trapping potential introduces an additional length scale $\ell_t$
into the problem, which drastically changes the general features of
the behavior at a phase transition.  For example, the correlation
functions of the critical modes do not develop a diverging length
scale in a finite trap.  Nevertheless, when the trap size $\ell_t$
becomes large, we may still observe a critical regime around the
transition point, with universal trap-size scaling (TSS) behaviors
with respect to the trap size $\ell_t$. TSS is controlled by the
universality class of the phase transition of the homogeneous system.
It has some analogies with the standard FSS for homogeneous systems
which we exploited in our numerical study, see Sec.~\ref{qmc}.  The
main difference is that, at the critical point, the correlation length
$\xi$ around the center of the trap shows a nontrivial power-law
dependence on the trap size $\ell_t$, i.e., $\xi\sim \ell_t^\theta$
where $\theta$ is the universal {\em trap} exponent.  TSS has been
numerically checked at the BEC transition of a single 3D bosonic
gas.\cite{CTV-13,CN-14} Analogous TSS arguments can be applied to the
BEC transitions of two-component bosonic gases.  In particular, the
trap exponent in the case of the harmonic space-dependence of $V_t$
turns out to be~\cite{CV-09}
\begin{equation}
\theta = {2\nu\over 1 + 2 \nu},
\label{theta}
\end{equation}
where $\nu$ is the correlation-length exponent of the transition, thus
$\theta=0.57327(4)$ at the decoupled XY fixed point controlling the
simultaneous condensation of both components, and $\theta\approx 0.5$
when only one component condenses.

Our study is relevant to experiments in which a mixture of two bosonic
atomic vapours is cooled to the point that at least one of them
undergoes a BEC transition and the number of particles is separately
conserved between the two species.  Recent years have seen the
development of many such experiments, with the two bosonic species
being two hyperfine levels of a single isotope
\cite{PhysRevLett.78.586, PhysRevLett.81.1539, PhysRevLett.82.2228,
  PhysRevLett.99.190402, PhysRevLett.103.245301,
  PhysRevLett.105.045303, PhysRevA.82.033609, PhysRevLett.85.2413,
  PhysRevA.63.051602, PhysRevA.80.023603}, two isotopes of the same
element \cite{PhysRevLett.101.040402, PhysRevA.84.011610,
  PhysRevA.79.021601} or heteronuclear mixtures of different elements
\cite{PhysRevLett.89.053202, PhysRevLett.89.190404,
  PhysRevLett.99.010403, PhysRevA.77.011603, PhysRevLett.100.210402}.
Notably, some mixtures were also successfully loaded on optical
lattices \cite{PhysRevLett.105.045303, PhysRevA.77.011603,
  PhysRevLett.100.210402}.  The availability of a wide range of atomic
species and the presence of Fano-Feshbach resonances allow to tune the
intra-species and inter-species interactions between the two
components of the mixtures.  Additional control can be achieved by
acting on the depth of an optical lattice. The high degree of
tunability of these systems may make the direct observation of the
transitions we predict within the reach of experiments.

\begin{acknowledgements}                                                      
We acknoledge discussions with D. Ciampini, L. Pollet  and S. Wessel.
We acknowledge computing time at the Scientific Computing Center of INFN-Pisa.
\end{acknowledgements}


\begin{thebibliography}{99}

\bibitem{CW-02} E.A. Cornell and C.E. Wieman,
Nobel Lecture: Bose-Einstein
  condensation in a dilute gas, the first 70 years and some recent
  experiments,
Rev. Mod. Phys. {\bf 74}, 875 (2002).

\bibitem{Ketterle-02} N. Ketterle, Nobel lecture: When atoms behave as
  waves: Bose-Einstein condensation and the atom laser,
  Rev. Mod. Phys. {\bf 74}, 1131 (2002).

\bibitem{BDZ-08}
I. Bloch, J. Dalibard, and W. Zwerger,
Many-body physics with ultracold gases,
Rev. Mod. Phys. {\bf 80}, 885 (2008).

\bibitem{Lipa-etal-96} J.A. Lipa, D.R. Swanson, J.A. Nissen,
  T.C.P. Chui, and U.E. Israelsson, Heat Capacity and Thermal Relaxation
  of Bulk Helium very near the Lambda Point, Phys. Rev. Lett. {\bf 76}, 944
  (1996).

\bibitem{CHPV-06} M. Campostrini, M. Hasenbusch, A. Pelissetto, and
  E. Vicari, Theoretical estimates of the critical exponents of the
  superfluid transition in $^4$He by lattice methods, Phys. Rev. B {\bf 74},
 144506 (2006).

\bibitem{PV-02} A. Pelissetto and E. Vicari, Critical phenomena and
  renormalization-group theory, Phys. Rep. {\bf 368}, 549 (2002).

\bibitem{DRBOKS-07} T. Donner, S. Ritter, T. Bourdel, A. \"Ottl,
  M. K\"ohl, and T. Esslinger, Critical Behavior of a Trapped Interacting
  Bose Gas, Science {\bf 315}, 1556 (2007).




\bibitem{PhysRevLett.78.586}
C. J. Myatt, E. A. Burt, R. W. Ghrist, E. A. Cornell and C. E. Wieman,
Production of Two Overlapping Bose-Einstein Condensates by Sympathetic Cooling,
Phys. Rev. Lett. {\bf 78}, 586 (1997).

\bibitem{PhysRevLett.81.1539}
D. S. Hall, M. R. Matthews, J. R. Ensher, C. E. Wieman and E. A. Cornell,
Dynamics of Component Separation in a Binary Mixture of Bose-Einstein Condensates,
Phys. Rev. Lett. {\bf 81}, 1539 (1998).

\bibitem{PhysRevLett.82.2228} 
H. Miesner, D. M. Stamper-Kurn,
  J. Stenger, S. Inouye, A. P. Chikkatur and W. Ketterle, Observation
  of Metastable States in Spinor Bose-Einstein Condensates,
  Phys. Rev. Lett. {\bf 82}, 2228 (1999).

\bibitem{PhysRevA.63.051602} G. Delannoy, S. G. Murdoch, V. Boyer,
  V. Josse, P. Bouyer and A. Aspect, Understanding the production of
  dual Bose-Einstein condensation with sympathetic cooling,
  Phys. Rev. A {\bf 63}, 051602 (2001).

\bibitem{PhysRevLett.99.190402} K. M. Mertes, J. W. Merrill,
  R. Carretero-Gonz\'alez, D. J. Frantzeskakis, P. G. Kevrekidis and
  D. S. Hall, Nonequilibrium Dynamics and Superfluid Ring Excitations
  in Binary Bose-Einstein Condensates, Phys. Rev. Lett. {\bf 99},
  190402 (2007).

\bibitem{PhysRevLett.103.245301}
D. M. Weld, P. Medley, H. Miyake, D. Hucul, D. E. Pritchard and W. Ketterle,
Spin Gradient Thermometry for Ultracold Atoms in Optical Lattices,
Phys. Rev. Lett. {\bf 103}, 245301 (2009).

\bibitem{PhysRevA.80.023603}
R. P. Anderson, C. Ticknor, A. I. Sidorov and B. V. Hall,
Spatially inhomogeneous phase evolution of a two-component Bose-Einstein condensate,
Phys. Rev. A {\bf 80}, 023603 (2009).

\bibitem{PhysRevLett.105.045303}
B. Gadway, D. Pertot, R. Reimann and D. Schneble,
Superfluidity of Interacting Bosonic Mixtures in Optical Lattices,
Phys. Rev. Lett. {\bf 105}, 045303 (2010).

\bibitem{PhysRevA.82.033609} 
S. Tojo, Y. Taguchi, Y. Masuyama,
  T. Hayashi, H. Saito and T. Hirano, Controlling phase separation of
  binary Bose-Einstein condensates via mixed-spin-channel Feshbach
  resonance, Phys. Rev. A {\bf 82}, 033609 (2010).

\bibitem{Nature.396.345} J. Stenger, S. Inouye, D. M. Stamper-Kurn,
  H.-J. Miesner, A. P. Chikkatur and W. Ketterle, Spin domains in
  ground-state Bose-Einstein condensates, Nature {\bf 396}, 345 (1998).

\bibitem{PhysRevLett.85.2413}
P. Maddaloni, M. Modugno, C. Fort, F. Minardi and M. Inguscio,
Collective Oscillations of Two Colliding Bose-Einstein Condensates,
Phys. Rev. Lett. {\bf 85}, 2413 (2000).

\bibitem{PhysRevLett.89.053202}
G. Ferrari, M. Inguscio, W. Jastrzebski, G. Modugno, G. Roati and A. Simoni,
Collisional Properties of Ultracold K-Rb Mixtures,
Phys. Rev. Lett. {\bf 89}, 053202 (2002).

\bibitem{PhysRevLett.89.190404}
G. Modugno, M. Modugno, F. Riboli, G. Roati and M. Inguscio,
Two Atomic Species Superfluid,
Phys. Rev. Lett. {\bf 89}, 190404 (2002).

\bibitem{PhysRevLett.99.010403}
G. Roati, M. Zaccanti, C. D'Errico, J. Catani, M. Modugno, A. Simoni, M. Inguscio and G. Modugno,
$^{39}\mathrm{K}$ Bose-Einstein Condensate with Tunable Interactions,
Phys. Rev. Lett. {\bf 99}, 010403 (2007).

\bibitem{PhysRevA.77.011603}
J. Catani, L. De Sarlo, G. Barontini, F. Minardi and M. Inguscio,
Degenerate Bose-Bose mixture in a three-dimensional optical lattice,
Phys. Rev. A {\bf 77}, 011603 (2008).

\bibitem{PhysRevLett.100.210402}
G. Thalhammer, G. Barontini, L. De Sarlo, J. Catani, F. Minardi and M. Inguscio,
Double Species Bose-Einstein Condensate with Tunable Interspecies Interactions,
Phys. Rev. Lett. {\bf 100}, 210402 (2008).

\bibitem{PhysRevLett.101.040402}
S. B. Papp, J. M. Pino and C. E. Wieman,
Tunable Miscibility in a Dual-Species Bose-Einstein Condensate,
Phys. Rev. Lett. {\bf 101}, 040402 (2008).

\bibitem{PhysRevA.79.021601}
T. Fukuhara, S. Sugawa, Y. Takasu and Y. Takahashi,
All-optical formation of quantum degenerate mixtures,
Phys. Rev. A {\bf 79}, 021601 (2009).


\bibitem{PhysRevA.84.011610}
S. Sugawa, R. Yamazaki, S. Taie and Y. Takahashi,
Bose-Einstein condensate in gases of rare atomic species,
Phys. Rev. A {\bf 84}, 011610 (2011).



\bibitem{HS-96}
T.-L. Ho and V.B. Shenoy,
Binary mixtures of Bose condensates of alkali atoms,
Phys. Rev. Lett. {\bf 77}, 3276 (1996).

\bibitem{Boninsegni-01} M. Boninsegni, Phase separation in mixtures of
  hard-core bosons, Phys. Rev. Lett. {\bf 87}, 087201 (2001).

\bibitem{AHDL-03} E. Altman, W. Hofstetter, E. Demler, and M.D. Lukin,
  Phase diagram of two-component bosons on an optical lattice, New
  J. Phys. {\bf 5}, 113 (2005).

\bibitem{DDL-03} L.-M. Duan, E. Demler, and M.D. Lukin, Controlling
  spin exchange interactions of ultracold atoms in optical lattices,
  Phys. Rev. Lett. {\bf 91}, 090402 (2003).

\bibitem{KS-03} A.B. Kuklov and B.V. Svistunov, Counterflow
  superfluidity of two-species ultracold atoms in a commensurate
  optical lattice, Phys. Rev. Lett. {\bf 90}, 100401 (2003).

\bibitem{PC-03} B. Paredes and J.I. Cirac, From Cooper pairs to
  Luttinger liquids with Bosonic atoms in optical lattices,
  Phys. Rev. Lett. {\bf 90}, 150402 (2003).

\bibitem{KG-04}
K.V. Krutitsky and R. Graham,
Spin-1 bosons with coupled ground states in optical lattices,
Phys. Rev. A {\bf 70}, 063610 (2004).

\bibitem{KPS-04} A.B. Kuklov, N. Prokof'ev, and B.V. Svistunov,
  Superfluid-superfluid phase transitions in a two-component
  Bose-Einstein condensate, Phys. Rev. Lett. {\bf 92}, 030403 (2004).

\bibitem{ICSG-05} A. Isacsson, M.-C. Cha, K. Sengupta, and
  S.M. Girvin, Superfluid-insulator transitions of two-species bosons
  in an optical lattice, Phys. Rev. B {\bf 72}, 184507 (2005).

\bibitem{PSP-08} R.V. Pai, K. Sheshadri and R. Pandit, Phases and
  transitions in the spin-1 Bose-Hubbard model: Systematics of a
  mean-field theory, Phys. Rev. B {\bf 77}, 014503 (2008).

\bibitem{SCPS-09} S.G. S\"oyler, B. Capogrosso-Sansone,
  N.V. Prokof'ev, and B.V. Svistunov, Sign-alternating interaction
  mediated by strongly correlated lattice bosons, New. J. Phys. {\bf
    11}, 073036 (2009).

\bibitem{HSH-09} A. Hubener, M. Snoek and W. Hofstetter, Magnetic
  phases of two-component ultracold bosons in an optical lattice,
  Phys. Rev. B {\bf 80}, 245109 (2009).


\bibitem{CSPS-10} B. Capogrosso-Sansone, S.G. S\"oyler,
  N.V. Prokof'ev, and B.V. Svistunov, Critical entropies for magnetic
  ordering in bosonic mixtures on a lattice, Phys. Rev. A {\bf 81},
  053622 (2010).

\bibitem{FHRSB-11} L. de Forges de Parny, F. H\'ebert, V.G. Rousseau,
  R.T. Scalettar, and G.G. Batrouni, Ground state phase diagram of
  spin-1/2 bosons in a two-dimensional optical lattice, Phys. Rev. B
  {\bf 84}, 064529 (2011).

\bibitem{Pollet-12} L. Pollet, Recent developments in Quantum
  Monte-Carlo simulations with applications for cold gases,
  Rep. Prog. Phys. {\bf 75}, 094501 (2012).

\bibitem{ACV-14} A. Angelone, M. Campostrini, and E. Vicari, Universal
  quantum behaviors of interacting fermions in 1D traps: from few
  particles to the trap thermodynamic limit, Phys. Rev.  A {\bf 89}, 023635
  (2014).

\bibitem{LCD-14} J.-P. Lv, Q.-H. Chen, and Y. Deng, Two-species
  hard-core bosons on the triangular lattice: A quantum Monte Carlo
  study, Phys. Rev. A {\bf 89}, 013628 (2014).

\bibitem{GBS-15} P.N. Galteland, E. Babaev, and A. Sudb\o, Immiscibile
  two-component Bose Einstein condensates beyond mean-field
  approximation: phase transitions and rotational response,
  arXiv:1503.05583.

\bibitem{CN-14} G. Ceccarelli and J. Nespolo, Universal scaling of
  three-dimensional bosonic gases in a trapping potential,
  Phys. Rev. B {\bf 89}, 054504 (2014).

\bibitem{CTV-13} G. Ceccarelli, C. Torrero, and E. Vicari, Critical
  parameters from trap-size scaling in trapped particle systems,
  Phys. Rev. B {\bf 87},  024513 (2013).

\bibitem{CR-12} J. Carrasquilla and M. Rigol, Superfluid to normal
  phase transition in strongly correlated bosons in two and three
  dimensions, Phys. Rev. A {\bf 86}, 043629 (2012).

\bibitem{CPS-07} B. Capogrosso-Sansone, N.V. Prokof'ev, and
  B.V. Svistunov, Phase diagram and thermodynamics of the
  three-dimensional Bose-Hubbard model, Phys. Rev. B {\bf 75}, 134302
  (2007).

\bibitem{Kawamura-88} H. Kawamura, 
Renormalization-group analysis of chiral transitions,
Phys. Rev. B {\bf 38}, 4916 (1988).

\bibitem{Aharony-76} A. Aharony, Dependence of universal critical
  behaviour on symmetry and range of interaction.  -- In: Phase
  Transitions and Critical Phenomena.  Vol. 6, edited by C. Domb and
  M.S. Green, (Academic, New York, 1976) p. 357.

\bibitem{ZJ-book} 
J. Zinn-Justin, {\em Quantum Field Theory and Critical Phenomena},
fourth edition (Clarendon Press, Oxford, 2002).

\bibitem{v-07} E. Vicari, Critical phenomena and renormalization-group
  flow of multi-parameter $\Phi^4$ field theories, PoS (LAT2007) 023;
  arXiv:0709.1014.

\bibitem{PV-05} A. Pelissetto and E. Vicari, Interacting $N$-vector order
  parameters with O($N$) symmetry, Condensed Matter Physics (Ukraine)
  {\bf 8}, 87 (2005).

\bibitem{PRV-01}
A. Pelissetto, P. Rossi, and E. Vicari,
The critical behavior of frustrated spin models with noncollinear order,
Phys. Rev. B {\bf 63}, 140414(R) (2001).

\bibitem{CPPV-04} P. Calabrese, P. Parruccini, A. Pelissetto,
  and E. Vicari, Critical behavior of O(2)$\otimes$O($N$)-symmetric
  models, Phys. Rev. B {\bf 70}, 174439 (2004).

\bibitem{foot-controversy} The existence of the chiral FP in
  O($N$)$\otimes$O(2) models for $N=2,3$ has been questioned in
  several papers.  See, e.g., M. Tissier, B. Delamotte, and
  D. Mouhanna, XY frustrated systems: continuous exponents in
  discontinuous phase transitions, Phys. Rev. B {\bf 67}, 134422
  (2003), for alternative interpretations of the theoretical and
  experimental results.  However, quite recently a chiral FP has been
  identified in the O(3)$\otimes$O(2) theory, using a completely
  different method, the so-called conformal bootstrap approach: see
  Y. Nakayama and T. Ohtsuki, Bootstrapping phase transitions in QCD
  and frustrated spin systems, Phys. Rev. D {\bf 91}, 021901 (2015).
  These results (although they only apply to $N=3$) further support
  the existence of a chiral FP in these models.

\bibitem{HV-11} M. Hasenbusch and E. Vicari, Anisotropic perturbations
  in three-dimensional O($N$)-symmetric vector models, Phys. Rev. B
  {\bf 84}, 125136 (2011).

\bibitem{Hasenbusch-01} M. Hasenbusch, Eliminating leading corrections
  to scaling in the three-dimensional $O(N)$-symmetric $\phi^4$ model: $N = 3$
  and 4, J. Phys. A {\bf 34}, 8221 (2001).

\bibitem{CPV-03} P. Calabrese, A. Pelissetto, and E. Vicari, Multicritical
  behavior of ${\rm O}(n_1)\oplus {\rm O}(n_2)$-symmetric systems, 
Phys. Rev. B {\bf 67}, 054505 (2003).

\bibitem{NKF-74}
D. R. Nelson, J. M. Kosterlitz, and M.i E. Fisher, 
Renormalization-Group Analysis of Bicritical and Tetracritical Points,
Phys. Rev. Lett.  {\bf 33}, 813 (1974).

\bibitem{KNF-76}
J. M. Kosterlitz, D. R. Nelson, and M. E. Fisher, 
Bicritical and tetracritical points in anisotropic antiferromagnetic systems,
Phys. Rev. B  {\bf 13}, 412  (1976).

\bibitem{Aharony-02} A. Aharony, Comment on “Bicritical and
  Tetracritical Phenomena and Scaling Properties of the SO(5) Theory”,
  Phys. Rev. Lett.  {\bf 88}, 059703 (2002).

\bibitem{1DHM}
F. H. L. Essler, H. Frahm, F. G\"{o}hmann, A. Kl\"{u}mper, 
and V. E. Korepin,
\emph{The One--Dimensional Hubbard Model}
(Cambridge University Press, Cambridge, 2005).

\bibitem{SK-91}
A. W. Sandvik and J. Kurlij\"arvi,
Quantum Monte Carlo simulation method for spin systems,
Phys. Rev. B {\bf 43}, 5950 (1991).

\bibitem{SS-02}
O. F. Sylju\aa{}sen and A. W. Sandvik,
Quantum Monte Carlo with directed loops,
Phys. Rev. E {\bf 66}, 046701 (2002).

\bibitem{DT-01} A. Dorneich and M. Troyer, Accessing the dynamics of
  large many-particle systems using the stochastic series expansion,
  Phys. Rev. E {\bf 64}, 066701 (2001).

\bibitem{footnote-mc} Our implementation of the directed operator-loop
  SSE algorithm is a natural extension to the two-species hard-core BH
  model of the single species algorithm.  There are 32 matrix
  elements, giving rise to $48$ possible transitions for each species.
  The operator loops always act on a single species at the time: at
  the beginning of each loop, we randomly select the species to
  update.


\bibitem{footnote_cr} At fixed $U$, for each $L$, we interpolated the
  data in a neighborhood of the crossing point with a cubic
  polynomial (or with a quadratic one if not enough data points are
  available for a cubic regression).  Subsequently, we numerically
  evaluated the crossing point between the interpolants, using
  bootstrap methods to estimate the error.

\bibitem{footnote_betac} The fit is repeated, progressively discarding
  the data from the smaller lattice sizes, i.e., by only fitting the
  crossing points for which $L_1 > L_{\min}$.  This allows us to
  control residual corrections to scaling.

\bibitem{Berezinski-72}
V. L. Berezinski,
Destruction of long-range order in one-dimensional and 2-dimensional systems
having a continuous symmetry group. 1. Classical systems,
Sov. Phys. JETP {\bf 32}, 493 (1971).

\bibitem{KT-73}
J. M. Kosterlitz and D. J. Thouless,
Ordering, metastability and phase transitions in two-dimensional systems,
J. Phys. C {\bf 6}, 1181 (1973).

\bibitem{Kosterlitz-74}
J. M. Kosterlitz,
Critical properties of the 2-dimensional XY model,
J. Phys. C {\bf 7}, 1046 (1974).

\bibitem{HPV-05}
M. Hasenbusch, A. Pelissetto, and E. Vicari,
Multicritical behavior in the fully frustrated XY model and related systems,
J. Stat. Mech. (2005) P12002.

\bibitem{AGG-80}
D. J. Amit, Y. Y. Goldschmidt, and G. Grinstein,
Renormalization group analysis of the phase transition in the 2D Coulomb
gas, sine-Gordon theory and XY model,
J. Phys. A {\bf 13}, 585 (1980).

\bibitem{PV-13}
A. Pelissetto and E. Vicari,
Renormalization-group flow and asymptotic behaviors at the
Berezinskii-Kosterlitz-Thouless transitions,
Phys. Rev. E {\bf 87}, 032105 (2013).

\bibitem{CNPV-13}
G. Ceccarelli, J. Nespolo, A. Pelissetto, and E. Vicari,
Universal behavior of two-dimensional bosonic gases at 
Berezinskii-Kosterlitz-Thouless 
transitions, Phys. Rev.  B  {\bf 88}, 024517 (2013).


\bibitem{CV-09} M. Campostrini and E. Vicari, Critical behavior and
  scaling in trapped systems, Phys.\ Rev.\ Lett.\ {\bf 102}, 240601
  (2009); (E) {\bf 103}, 269901 (2009); Trap-size scaling in confined
  particle systems at quantum transitions, Phys. Rev. A {\bf 81},
  023606 2010.

\end{thebibliography}
\end{document}